\documentclass[11pt,a4paper]{article}
\usepackage{version}
\usepackage{float}
\usepackage{amsmath}
\usepackage{amstext}
\usepackage{tabularx}
\usepackage[integrals]{wasysym}
\usepackage{marvosym}
\usepackage{manfnt}
\usepackage{ifsym}
\usepackage[psamsfonts]{amssymb}
\usepackage[unicode,bookmarks,bookmarksopen,bookmarksopenlevel=0,colorlinks,%
plainpages=false,pdfpagelabels,hypertexnames=false,pageanchor=true,breaklinks=true,%
linkcolor=blue,citecolor=blue,urlcolor=red]{hyperref}
\usepackage[left=1.5cm, right=1.5cm, top=1.0cm, bottom=1.5cm,
includehead]{geometry}
\usepackage{upref}  %ссылки независимо от контекста печатаются прямыми %
\usepackage{indentfirst}
\newif\ifpdf
\ifx\pdfoutput\undefined
\pdffalse % we are not running PDFLaTeX
\else
\pdfoutput=1 % we are running PDFLaTeX
\pdftrue \fi
\ifpdf
\usepackage[pdftex]{graphicx}
\usepackage{graphicx}
\else
\usepackage{graphicx}
\fi

\newcommand\pr{\partial}
\newcommand\ben{\begin{enumerate}}
\newcommand\een{\end{enumerate}}
\newcommand\bed{\begin{itemize}}
\newcommand\eed{\end{itemize}}
\newcommand\bei{\begin{description}}
\newcommand\eei{\end{description}}
\newcommand\be{\begin{equation}}
\newcommand\ee{\end{equation}}
\newcommand\beq{\begin{eqnarray}}
\newcommand\eeq{\end{eqnarray}}
\newcommand\beqo{\begin{eqnarray*}}
\newcommand\eeqo{\end{eqnarray*}}

\newtheorem{lemma}{Lemma }%[section]
\newtheorem{theorem}{Theorem}%[section]
\newtheorem{rmrk}{Remark}

\newcommand\IM{\operatorname{Im} \,}

\newcommand\const{\operatorname{const}}

\def\be{{\bf e}}

\newcommand{\byd}{\stackrel{\mathrm{def}}{=}}

\usepackage{fancyhdr}
\usepackage{lastpage}
\begin{document}
\title{ {\small PREY-TAXIS VS AN EXTERNAL SIGNAL:}\\ {\small SHORT-WAVE ASYMPTOTIC AND STABILITY ANALYSIS}}
\author{{\small Andrey Morgulis\footnote{ORCID 0000-0001-8575-4917X, abmorgulis@sfedu.ru}}, \\
{\footnotesize I.I.Vorovich Institute for Mathematic, Mechanics and Computer Science,}\\
{\footnotesize Southern Federal University, Rostov-na-Donu, Russia;\quad}\\
{\footnotesize Southern Mathematical Institute of VSC RAS, Vladikavkaz, Russia}
 \and 
  {\small Karrar H. Mallal}\\
 {\footnotesize I.I.Vorovich Institute for Mathematics, Mechanics and Computer Sciences,}\\
{\footnotesize Southern Federal University, Rostov-na-Donu, Russia} \\
  }
%\vspace{3mm}\\
%\date{{\small 9,2024}}
\maketitle
% Remove command to get current date
\begin{abstract}
\noindent
  {\footnotesize We consider two models of predator-prey community with prey-taxis, one relies on Patlak-Keller-Segel law (Lee et al, 2009), the other one employs Cattaneo’s model of heat transfer following  Dolak \& Hillen (2003). Thus, the former one  uses the prey density gradient for directing  the predators flux, and the latter one -- for directing the  vector density of sources of the predators flux.   We  assume  the predators to be  also capable of responding to an external signal in the same manner.  Additionally, we assume  that some scaling  makes the dimensionless prey diffusivity and the wavelength of the external signal small quantities of the same order. In the case of Cattaneo's  model, we additionally assume the resistivity to varying the predators flux to be as high as the reciprocal of the prey diffusivity.  With these assumptions we construct the complete  asymptotic expansion of the    short-wave solution.  We use this asymptotic  to examine the effect of the short-wave signal on the formation of spatiotemporal patterns.  We do so  by comparing the stability of equilibria with no signal to that of the quasi-equilibria, by which we mean the simplest patterns directly forced by the external signal. Historically, such an approach goes back to Kapitza's theory for upside-down pendulum, and it had been already applied to examining other predator-prey systems  and/or  other  limits  (Morgulis \& Ilin, 2020, Morgulis, 2023). This time the overall conclusion is that  the external signal is likely not capable of creating  the instability domain in the parametric space from nothing but it can substantially  widen the one that is non-empty with no signal. The details, however, essentially depend on the speed at which the external signal propagates and on the system kinetics, at least when the signal amplitude is small.}
\end{abstract}
\section*{Introduction}\label{ScIntr}
\addcontentsline{toc}{section}{Introduction}
\noindent
We start with the explanation of several basic notions for helping a reader  who  is  not very much involved in our topic.  Consider a media consisting of the particles  that are  sensitive to several signals or stimuli in the sense that the media responds to the intensity gradient of a signal by  the directed movement that is  often called taxis. In particular, calling the name of chemotaxis supposes the signal to be a concentration of some chemical. In the mathematical literature, this name is in use for calling  the  models relying on the Patlak-Keller-Segel law that describes the media response by its  flux   directed along  the  signal gradient. Further, such name as `the Keller-Segel chemotaxis' often presumes additionally a specific  kinetics of the chemical.  The intensive studying of this class of  system  goes back for several decades.  To see the main outcomes,  a reader can refer to the reviews \cite{Hrstmnn,HllnPntr,BellBellTao}.  

Meanwhile, article \cite{KrvOdll} had reported the prey-taxis phenomenon, and the area of applications of PKS law had  extended to  modelling the spatially distributed predator-prey communities \cite{Ts94}, \cite{BrzKrv}. Later on, Lee, Hillen \& Lewis \cite{LeeHlnLws}  have systematically studied  the instabilities of the coexistence equilibrium in the parabolic predator-prey system with the prey-taxis directly expressed by PKS law. They revealed the substantial role  of the  system kinetics. For example, in the case of  Lotka-Volterra' kinetics, the coexistence equilibrium  turned out to  remain stable as long as it exists. In contrast, the Arditi-Ginzburg ratio-dependent kinetics \cite{ArG} leads to instabilities. Banerjee \& Petrovskii \cite{BnPtrRtDpnd} studied the  resulting patterns in more detail.

Some early studies  addressed the more complex variants of the tactical behaviour, e.g. pursuing-evasion scheme \cite{Ts04,Ts04-1} or the prey-induced acceleration \cite{GMT,AGMTS}. In the latter setting,  the effect of the prey-taxis on the coexistence predator-prey equilibrium turned out to be destabilizing even for the Lotka-Volterra kinetics. One more feature is that the mode of instability is presumable oscillatory  and inhomogeneous.  Tyutyunov et al. \cite{SapTA}, \cite{TtnZgr}, and Chakraborti et al. \cite{Chkr} have found  persisting the destabilizing effect of the prey-induced acceleration for several classes of  kinetics.  In the course of analysis, Tytyunov et al. have pointed out the equivalence between the simplest predator-prey systems with the indirect prey-taxis \cite{TllWrzk} and prey-induced acceleration. 

It worths noting here that  the  name  `prey-induced acceleration'' has appeared in recent article \cite{TaoW} instead of the original name,  `slow taxis, which  has been chosen to emphasize the capability of capturing the inertia of predators' orientation.  The inertial response brings the slow taxis or, equivalently, prey-induced acceleration  model  at a close relation to indirect prey-taxis,    which  means searching the prey by a signal they produce. The same reasoning links the  prey-induced acceleration to the hyperbolic models \cite{BellBellTao,Eft}, e.g. Cattaneo model for the chemosensitive movement \cite{Hillen},  that considers directing the sources density for predators' flux by the chemicals gradient.

Further, we by no means  have to neglect the case of sensitivity to more than one signal. The challenging examples arise from  chemotaxis- haptotaxis  (see e.g. the cited above review by Bellomo at al. or  article \cite{CrrtPrtVchlt}) or  the multi-level food chains \cite{Grchv}. When some of signals perceiving by the community are independent of its state, we consider them as the external ones. The need in addressing such signals arises, e.g.,  if we aim at taking into account the effects exerted from the environment. Moreover, some of the signals  can have very different time or\and space scales. For instance,  several authors have recently addressed a two time scale predator-prey model \cite{Chdh1,Chdh2,Chdh3}. In these papers, in particular, they discuss the biological grounds for considering the multiple scale models. 

Morgulis \& Ilin \cite{AM1} and Morgulis  \cite{AM2} addressed  the  predator-prey model with the prey-taxis and short-wave external signal. They considered both the parabolic formulation with the indirect prey-taxis (=prey driven acceleration=slow taxis)  and hyperbolic formulation of Cattaneo' kind. They applied the two scale expansions technique, see, e.g., \cite{Allr-1}, to derive the leading homogenized system, which they employed to examine the effect of the external signal.   Here we mean comparing the stability of the equilibria of the original system with no signal to those of the leading homogenized system. The latter ones actually describes the so-called quasi-equilibria -- that is, the simplest short wave patterns straightforwardly froced  by the signal.   This  approach goes back to upside-down pendulum \cite{LndLfs},  and today   it   is widely recognized that the short-wave/high-frequency forcing can  change drastically the area of stability of a quasi-equilibrium in the space of the system parameters compared to the equilibrium, or even change stability to instability and vice-versa.  Such effects  are interesting  for the population or cell dynamics given the common understanding of the importance of the instabilities and  related bifurcations for the spatio-temporal patterns formation.  In the mentioned articles, the authors performed the outlined analysis for Lotka-Volterra' kinetics.  It turned out that,  given the indirect taxis, the effect of the signal can be both stabilizing and destabilizing depending on the speed at which it propagates.  In the case of Cattaneo's model the effect is always destabilizing. It worths noting here, that, in both cases,  the equilibrium of the coexisting species undergo oscillatory instability with no external signal  too, in contrast with the model with the direct prey-taxis in the classical PKS formulation, where the same equilibrium remains stable as long as it exists (see the cited above article by Lee et al.).  

In the present article  we proceed to the direct prey-taxis   in the common parabolic PKS formulation, and also consider  Cattaneo' prey-taxis, when the prey density gradient directs the density of sources of the predators flux. We consider these models with both  Lotka-Volterrá's and ratio-depended  Arditi-Ginzbourg's kinetics, and implement the program outlined in the previous paragraph for all the four cases. 

%Additionally, we revisited the Cattaneo's model and found another asymptotic solution additionally assuming that varying predatos' flux in response to the prey-taxis encounters very high resistivity. This asymptotic seems less singular and more feasible to rigorous justifying than that built in the cited above article by Morgulis (2023).  We have considered Lotka-Volterra's and  Ginzbourg-Arditi's kinetics and found that the coexistence quasi-equilibria remains stable as long as it exists in the Lotka-Volterra's case, and destabilizes  in the  Ginzbourg-Arditi case.   The latter possibility seems unexpected amid the high resistivity, that stabilizes   the  coexistence quasi-equlibrium to a great extent with no external signal.   

The article is organized as follows. In Sec.~\ref{ScSttng} we formally set the models and introduce auxiliary matters. In Sec.~\ref{ScPKS} we derive the asymptotic solution for the  PKS prey-taxis.  In Sec.~\ref{ScCttn} we do the same for Cattaneo's  prey-taxis. In Sec.~\ref{ScStQsEq} we introduce the quasi-equilibria and perform the linear stability analysis. In Sec.~\ref{ScDscsn} we discuss the obtained results. Appendix is to expose the scaling adopted here.
\section{General settings}
\noindent
\subsection{Formulation of the problem}
\label{ScSttng}
\noindent 
We address two one-dimensional models  that read as
\begin{eqnarray}
 &p_t+{q}_x=pf(p,s),&
 \label{PrdTrnsEq}\\
  &s_t=Ds_{xx}+sg(p,s), &
\label{PryDnstEq}  
\end{eqnarray}
and either 
\begin{eqnarray}
  &{q}=p\left(\chi(s)s_x+\varkappa(s)h_x\right) -\mu(s) p_x.&
  \label{PKSFlxEq}
\end{eqnarray}
or 
\begin{eqnarray}
  &{q}_t+\nu {q}=p\left(\chi(s)s_x+\varkappa(s)h_x\right) -\mu(s) p_x&
  \label{CttnFlxEq}
\end{eqnarray}  
Here {$t$} is time, {$x\in\mathbb{R}$} is spatial coordinate, $\nu=\const>0$, $p=p(x,t)$, ${ q}={q}(x,t)$ -- density and flux of the predators.  We put the flux equation either the PKS \eqref{PKSFlxEq} or Cattaneo's way \eqref{CttnFlxEq} with the signals represented by functions  $s=s(x,t)$ and $h=h(x,t)$. We interpret the former one as prey' density and the latter one is external.  Coefficients  $\chi$ and $\kappa$ measure predators' sensitivity  to the signals. Coefficient $\mu$ measures predators diffusivity. Term $\nu q$ in the Cattaneo flux equation  equations  expresses the contributions from the resistance of the environment to the particles motion. Reproduction and diffusion of the prey is described by equation \eqref{PryDnstEq}, where $D=\const>0$ is the diffusion coefficient. In this way, equations \eqref{PrdTrnsEq},\eqref{PryDnstEq} and \eqref{PKSFlxEq} constitute the parabolic  Patlak-Keller-Siegel's type model or, briefly, PKS-model, and system \eqref{PrdTrnsEq},\eqref{PryDnstEq} and \eqref{CttnFlxEq} constitute the hyperbolic  Cattaneo's type model.  Regarding  both  models  we assume that they are in  the dimensionless form (see Appendix).

We  address the short travelling waves driven by an external signal, which itself is a short travelling wave, so that we set
\begin{equation}\label{ShrtWvSgnl}
  h=h(x,t,\eta), \eta=(x-ct)/\delta, \quad  h(x,t,\eta+2\pi)=h(x,t,\eta) \ \forall\, x,t,\eta.
\end{equation}
where parameter $\delta$ is small. Regarding the parameters of the governing equations we assume that they do not depend on $\delta$ except for  $D$ and $\nu$. Regarding them we assume the following
\begin{equation}\label{LmtBhvrPrmtr}
  D=\delta, \quad \nu=\bar{\nu}/\delta
\end{equation}
We'll be seeking for the power asymptotic having in the form 
\begin{equation}\label{AsmpExpnsn}
  (p,q,s)=\sum\limits_{k=0}^{K-1}({p}_k,q_k,s_k)(x,t,\eta)\delta^k+O(\delta^K),\ \delta\to+0,\ K\in \mathbb{N}.
\end{equation}
where all the coefficients are $2\pi-$periodic in $\eta$.   This  constraint  plays an  essential part in the sequel. In addition, we assume that
$$
\chi\equiv\const,\quad \kappa\equiv\const,\quad \mu\equiv\const 
$$
--- that is, the transport coefficients are constant from now on. Removing this constraint  very likely  requires no new ideas unless the transport coefficients also depend on the predators' density. Nevertheless, we put the general case out  the score of this article to avoid hiding the key points of the proposed analysis   among  cumbersome calculations. Besides, the case of constant transport coefficients  itself is of interest  and  anyway useful as a benchmark for studying the case of signal-dependency. 
\subsection{Auxiliary matters}
\label{ScAuxSttng}
\noindent
For $z\in \mathbb{C}\setminus\{\mathrm{Re}\,z=0\}$ consider  
the convolution 
\begin{equation}\label{Inv(D+z)}
(\mathrm{R}_zu)(\eta)=\mp\int\limits_{\xi>0} u(\eta\pm \xi)\exp(\pm z\xi)d\xi, 
\end{equation}
where the sign of $+$ ($-$) must be in both places while writing the integrands if $\mathrm{Re}\,z<0$ ($\mathrm{Re}\,z>0$), and  the opposite sign must appear before the integral.  Transform $u\mapsto \mathrm{R}_zu$  determines a  bounded linear operator  $\mathrm{L}_2(\mathbb{S})\to  \mathrm{L}_2(\mathbb{S})$,  where the notation of $\mathbb{S}$ stands for the standard unit circumference. Clearly, 
$$
\mathrm{R}_z=(z+\pr)^{-1}.
$$
The Fourier  representation brings this resolvent into the  same form in both semi-planes of variable $z$, namely,
\begin{equation}\label{Inv(D+z)Frr}
(\mathrm{R}_zu)(\eta)=\sum\limits_{k\in \mathbb{Z}} \frac{\hat{u}_k\mathrm{e}^{ik\eta}}{ik+z}. 
\end{equation}
where the notation of $\hat{u}_k$ stands for the Fourier coefficients of function $u$.  From here, it follows that operator-valued function  $z\mapsto \mathrm{R}_z$ is analytic in the punctured complex plain when  the punctures constitute the set $i\mathbb{Z}$, and its  restriction on subspace 
\begin{equation}\label{SbsTldL2S}
  {\widetilde{\mathrm{L}}}_2\left(\mathbb{S}\right)\byd \{u\in \mathrm{L}_2(\mathbb{S}):\, \hat{u}_0=0.
  \}
\end{equation}
allows analytic continuation to a neighbourhood of the origin.  Operator $\mathrm{R}_0=\lim_{z\to 0} \mathrm{R}_z:\widetilde{\mathrm{L}}_2\left(\mathbb{S}\right)\to \widetilde{\mathrm{L}}_2 \left(\mathbb{S}\right)$ represents the inverse to operator $\pr$ in $\widetilde{\mathrm{L}}_2\left(\mathbb{S}\right)$ and the left inverse in ${\mathrm{L}}_2\left(\mathbb{S}\right)$.  Furhter, 
\begin{equation}\label{InvD(z+D)}
\mathrm{G}_z\byd \mathrm{R}_z\mathrm{R}_0=\left(\pr(z+\pr)\right)^{-1}:\widetilde{\mathrm{L}}_2\left(\mathbb{S}\right)\to \widetilde{\mathrm{L}}_2\left(\mathbb{S}\right),
\end{equation}
 and Fourier' representation of operator $\mathrm{G}_z$ reads as
\begin{equation}\label{InvD(D+z)Frr}
(\mathrm{G}_zu)(\eta)\byd \sum\limits_{k\in \mathbb{Z}\setminus\{0\}} \frac{\hat{u}_k\mathrm{e}^{ik\eta}}{ik(z+ik)}.
\end{equation} 
We'll be using the averaging -- that is, evaluating the averaged value (if any) that reads as  
$$
\langle f  \rangle=\lim\limits_{L\to\infty}\frac{1}{2L}\int\limits_{-L}^{L}f(\eta)\,d\eta.
$$
The averaged value exists for every periodic function, and it is nothing else than its Fourier coefficient $\hat{f}_0$.  We assume the external signal to vansih on average -- that is,  we set 
\begin{equation}\label{SgnlVnshonAvrg}
 \langle h(x,t,\cdot)\rangle=0\ \forall x,t
\end{equation}
%\begin{equation}\label{LmtBhvrPrmtr1}
 % \kappa={\kappa}(s)\delta, \quad \mu={\mu}(s)\delta.
%\end{equation}
\section{ Asymptotic for the PKS prey-taxis}
\label{ScPKS}
\noindent
Here we consider the system consisting of equations \eqref{PrdTrnsEq},\eqref{PryDnstEq} and \eqref{PKSFlxEq}, where the external signal is the short travelling wave  given by expression  \eqref{ShrtWvSgnl}.  We'll be developing  the  asymptotic of the response assuming equalities  \eqref{LmtBhvrPrmtr} to hold true. More precisely, we only need in this section  the one that regards parameter $D$. 

The system under consideration written with the use of the fast variable reads as
\begin{eqnarray}
&\left(q-cp\right)^\prime=\delta\left(pf(p,s)-p_{t}-q_{x}\right),&
\label{PrdTrnsEqAsmptPKS}\\
&\mu p^\prime  -p(\kappa  h^\prime +\chi {s}^\prime) =\delta\left(p\left(\chi s_x+\kappa h_x\right) -\mu p_x-q\right),&
\label{FlxEqAsmptPKS}\\
&\left(c{s}  +  {s}^\prime\right)^\prime=-\delta\left(sg(p,s)+\delta s_{xx}+2s^\prime_x-s_t\right),&
\label{PryDnstEqAsmptPKS}
%\\
%&\tilde{s}=s-s_0=\delta s_1+\ldots,\ \bar{s}=s_0,&
%\label{StlSbr}
 \end{eqnarray}
where the notation of  $(\cdot)^\prime$ stands for derivative in the fast variable, $\eta$. Replacing unknowns $p,q,s$ by expansions \eqref{AsmpExpnsn} leads to a chain of equations  for the  sequence of  coefficients $p_k,q_k,s_k$, $k=0,1,2,\ldots$. For the compactness of layout, we'll be replacing by a notation of $\mathrm{Op}_k$, $k=1,2,\dots$,  any collections of those terms that operate neither the coefficients  $p_k,q_k,s_k$ nor their derivatives, but only the ones preceding them. For  $k\le 0$, $\mathrm{Op}_k=0$ by definition. 

%Also, for  function $y(s)$ the notation of $y^{(k)}$  will be standing for $k-$th Taylor's coefficient relative to $s=s_0$.

Thus,  we arrive at the chain of equations that read as  
\begin{eqnarray}
&\left(q_k-cp_k\right)^\prime=P_{k-1}-p_{k-1,t}-q_{k-1,x}+\mathrm{Op}_{k-1},&
\label{PrdTrnsEqChnPKS}\\
&\left(\frac{p_k}{E}\right)^\prime 
-\frac{{\chi}p_0s^\prime_k}{{\mu}E} = \frac{{\chi}(p_0s_{k-1,x}+p_{k-1}s_{0,x})+\kappa p_{k-1}h_x-{\mu}p_{k-1,x}-q_{k-1}+\mathrm{Op}_{k-1}}{{\mu}E},&
\label{FlxEqChnPKS}\\
&\left(c{s}_k  +  {s}_k^\prime\right)^\prime=s_{k-1,t}-S_{k-1}-2s^\prime_{k-1,x}+\mathrm{Op}_{k-1},&
\label{PryDnstEqChnPKS}\\
&k=0,1,\ldots,&\nonumber
 \end{eqnarray}
where all the quantities bearing negative  indices are equal to zero,   terms $P_{k-1}$,  and $S_{k-1}$ are linear functions  in variables $p_{k-1}$,  $s_{k-1}$,   with coefficients depending on functions $s_0$ and $p_0$, and 
$$
E=E(x,t,\eta)=\exp\left(\frac{{\kappa}\mathrm{R}_0{h}^\prime+{\chi} \mathrm{R}_0 s_0^\prime}{{\mu}}\right).
$$
Chain of Eqs.~\eqref{PrdTrnsEqChnPKS}-\eqref{PryDnstEqChnPKS} follows from the Eqs.~\eqref{PrdTrnsEqAsmptPKS}-\eqref{PryDnstEqAsmptPKS}  by zeroing the collection of terms of order $\delta^{k}, k=0,1,\ldots$.  Deriving  Eq.~\eqref{FlxEqChnPKS} also involves  a simple transformation,  which we omit. 

For $k=0$, Eqs.~\eqref{PrdTrnsEqChnPKS}-\eqref{PryDnstEqChnPKS} are homogeneous in variables $p_0,q_0,s_0$ and easy to solve.   So, the leading term in asymptotic \eqref{AsmpExpnsn} reads as
\begin{equation}\label{P0Q0S0-PKS}
  s_0=\bar{s}_0(x,t),\quad p_0=\bar{r}_0(x,t)E(x,t,\eta),\quad q_0=cp_0(x,t,\eta)+\bar{v}_0(x,t),\ E=\mathrm{e}^{\frac{{\kappa}\mathrm{R}_0{h}^\prime}{{\mu}}},
\end{equation}
where functions in the slow variables, $x,t$, denoted here as $\bar{s}_0$, $\bar{v}_0$ and $\bar{r}_0$ play the part of constants of the integration in the fast variable. They are unknown at this stage. 

Given equalities~\eqref{P0Q0S0-PKS}, we turn to   system  Eqs.~\eqref{PrdTrnsEqChnPKS}-\eqref{PryDnstEqChnPKS} for $k=1,2,\ldots$. 
For resolving it  in the functions periodic in the fast variable,  there is a triad of solvability condition to hold true, namely 
\begin{eqnarray}
&\left\langle P_{k-1}-p_{k-1,t}-q_{k-1,x}+\mathrm{Op}_{k-1}\right\rangle=0 ,&
\label{PrdTrnsSlvChnPKS}\\
& \left\langle{\chi}\bar{r}_0 s_{k-1,x} +\frac{p_{k-1}({\chi}s_{0,x}+\kappa h_x)-{\mu}p_{k-1,x}-q_{k-1}+\mathrm{Op}_{k-1}}{E}\right\rangle=0,&
\label{FlxSlvChnPKS}\\
&\left\langle s_{k-1,t}-S_{k-1}+\mathrm{Op}_{k-1}\right\rangle=0.&
\label{PryDnstSlvChnPKS}
 \end{eqnarray}
It worths noting here that ${\chi}p_0/E={\chi}\bar{r}_0$ by equality \eqref{P0Q0S0-PKS}.

Assume these conditions  being met. Then we get the solutions that read as 
\begin{eqnarray}
&p_k =\bar{r}_k E+E\mathrm{ R}_0\left(\frac{{\chi}\bar{r}_0 ( s^\prime_k +  s_{k-1,x})}{{\mu}}+ \frac{p_{k-1}({\chi}s_{0,x}+\kappa h_x)-{\mu}p_{k-1,x}-q_{k-1}+\mathrm{Op}_{k-1}}{{\mu}E}\right),&
\label{Pk-PKS}\\
&q_k=\bar{v}_k+cp_k+\mathrm{R}_0(P_{k-1}-p_{k-1,t}-q_{k-1,x}+\mathrm{Op}_{k-1}),&
\label{Qk-PKS}\\
&{s}_k=\bar{s}_k+\mathrm{G}_c\left(s_{k-1,t}-S_{k-1}-2s^\prime_{k-1,x}+\mathrm{Op}_{k-1}\right),&
\label{Sk-PKS}
 \end{eqnarray}
where the constants of integration, $\bar{r}_k,\bar{v}_k,\bar{s}_k$ are unknown functions in the slow variables only. For what follows, it's important that  
$$
{s}_k-\bar{s}_k=\mathrm{Op}_k,\quad q_k-\bar{v}_k-cp_k=\mathrm{Op}_k,\quad {p_k}{E}^{-1}-\bar{r}_k=\mathrm{Op}_k,
$$
where all three right hand sides are zero on average.  This observation can appear  not completely obvious since there is  the term in Eq.~\eqref{Pk-PKS} involving $s^\prime_k$. To get rid of it, note that
$$
\mathrm{ R}_0 s^\prime_k={s}_k-\bar{s}_k=\mathrm{G}_c\left(s_{k-1,t}-S_{k-1}-2s^\prime_{k-1,x}+\mathrm{Op}_{k-1}\right).
$$
The endpoint expression is evidently $\mathrm{Op}_k$, and it is zero on average by definition of operator $\mathrm{G}_c$ (see the end of Section~\ref{ScAuxSttng}).  

Further,   we use the unknown slow functions to meet the solvability conditions for the next iteration. Eliminating  the $k-$th order corrections from there by virtue of Eqs.~\eqref{Pk-PKS}-\eqref{Sk-PKS}, we arrive at the equations for the slow functions, which read as  
\begin{eqnarray}
& \bar{p}_{k,t}+\bar{q}_{k,x}=\left\langle P_{k}+\mathrm{Op}_{k}\right\rangle,&
\label{PrdTrnsHmgnzdChnPKS}\\
& \bar{q}_{k} =\bar{\chi}\bar{p}_0\bar{s}_{k,x}+(\bar{\chi}\bar{s}_{0,x}+\bar{c})\bar{p}_k-{\bar{\mu}}\bar{p}_{k,x}+\langle\mathrm{Op}_k\rangle, 
&
\label{FlxHmgnzdChnPKS}\\
& \bar{s}_{k,t}-\left\langle S_{k}+\mathrm{Op}_{k}\right\rangle=0,&
\label{PryDnstHmgnzdChnPKS}\\
&P_k=a_{11}p_k+a_{12}s_k,\quad S_k=a_{21}p_k+a_{22}s_k,\qquad k=1,2,\ldots,&
\label{PkSk}\\
&a_{11}=f(p_0,s_0)+p_0f_p(p_0,s_0),\  a_{12}=p_0f_s(p_0,s_0),\ a_{21}=s_0g_p(p_0,s_0),\   a_{22}=g(p_0,s_0)+s_0g_s(p_0,s_0),&
\label{aij}\\
&\bar{\mu}=\frac{\mu } {\langle E^{-1}\rangle\langle E\rangle},\quad \bar{\chi}=\frac{\mu } {\langle E^{-1}\rangle\langle E\rangle},  \quad \bar{c}=c\left(1-\frac{1 } {\langle E^{-1}\rangle\langle E\rangle}\right)+\mu\left(\ln\,\langle E\rangle\right)_x.&
\label{BrCffcts}
 \end{eqnarray}
 where formula \eqref{P0Q0S0-PKS} determines function $E$. 
The linkage of dependent variables, $\bar{p}_k$, $\bar{q}_k$, to the constant of integration, $\bar{r}_k,\bar{v}_k$, and to the averaged values of the current coefficients follows from Eqs.~\eqref{Pk-PKS}-\eqref{Qk-PKS}, and we stress the following relations
 \begin{eqnarray}
&\langle {p}_k\rangle\byd \bar{p}_k =\bar{r}_k \langle E\rangle+\langle E\mathrm{R}_0 \mathrm{Op}_k\rangle,\quad \langle q_k\rangle\byd \bar{q}_k=\bar{v}_k+c\bar{p}_k,\quad \langle{s}_k\rangle=\bar{s}_k,\quad k=0,1,\ldots.&
\label{BrPkBrQkBrSk-PKS}
 \end{eqnarray}
 We call system \eqref{PrdTrnsHmgnzdChnPKS}-\eqref{PryDnstHmgnzdChnPKS} \emph{homogenized or slow system of order $k$, k=1,2,..}. The \emph{leading homogenized or slow  system} corresponds to $k=0$. We had better  formulate it separately because it does not arise from equations \eqref{PrdTrnsHmgnzdChnPKS}-\eqref{PryDnstHmgnzdChnPKS} just by setting $k=0$. 

Thus, developing the asymptotic expansion \eqref{AsmpExpnsn}    goes on iteration by iteration while we manage to solve the homogenized systems  starting from the leading one, which reads as
\begin{eqnarray}
& \bar{p}_{t}+\bar{q}_{x}= \bar{p}\bar{f}(\bar{p},\bar{s} ),&
\label{PrdTrnsHmgnzdLdPKS}\\
& \bar{q}=\bar{p}\left(\bar{\chi}\bar{s}_{x}+\bar{c}\right)-\bar {\mu}\bar{p}_{x},&
\label{FlxHmgnzdLdPKS}\\
& \bar{s}_t=\bar{s}\bar{g}(\bar{p},\bar{s} ), &
\label{PryDnstHmgnzdLdPKS}\\
&\bar{f}(\bar{p},\bar{s} )=\langle \bar{E}f(\bar{p}\bar{E},\bar{s})\rangle,\quad \bar{g}(\bar{p},\bar{s} )=\langle{g}(\bar{p}\bar{E},\bar{s} )\rangle,\ \bar{E}=\frac{E}{\langle E\rangle},&
\label{BrFBrGPKS}
 \end{eqnarray}
where Eq.~\eqref{P0Q0S0-PKS} determines  function $E$, and  we have omitted the lower indices.  Indeed, assume a solution to the leading slow system to be known. Then put explicitly  Eqs.~\eqref{BrPkBrQkBrSk-PKS} where $k=0$ and find auxiliary functions $\bar{r}_0$ and $\bar{v}_0$. Next determine the leading term of asymptotic expansion \eqref{AsmpExpnsn}, i.e., functions $p_0,q_0,s_0$, by Eqs.~(\ref{P0Q0S0-PKS}). 

Further, knowing the leading approximation,  put  explicitly  the first order slow system and system \eqref{BrPkBrQkBrSk-PKS} with $k=1$.   Solve the former one, and then use the solution,  $\bar{p}_1,\bar{q}_1,\bar{s}_1$, to determine functions $\bar{r}_1$ and $\bar{v}_1$ from the latter one. Next   define the first order terms of asymptotic expansion \eqref{AsmpExpnsn}, $p_1,q_1,s_1$,  by Eqs.~\eqref{Pk-PKS}-\eqref{Sk-PKS}.  

Further, knowing the zero and the first order terms allows us to write explicitly the second order slow system and  system \eqref{Pk-PKS}-\eqref{Sk-PKS} for $k=2$. Solving the former allows us to write explicitly  formulae Eq.~\ref{BrPkBrQkBrSk-PKS} with $k=2$. Thus we can get second order terms in the same manner as the first order ones. Obviously, we can  get in this  way as many terms of the asymptotic  expansion \eqref{AsmpExpnsn} as we wish.
\begin{rmrk}\label{RmLdSlwSstPKS}
 {\rm  It looks like   a common parabolic PKS-model  except for the \emph{drift} -- that is, the additional  flux, $\bar{c}\bar{p}$, where formula \eqref{BrCffcts} specifies the drift speed, $\bar{c}$.  Additionally, the transport coefficients undergo re-scaling also determined by  formula \eqref{BrCffcts}. The kinetic terms generally undergo changing too  as formula \eqref{BrFBrGPKS} shows, but with the notable exception for  the Lotka-Volterra case (see Sec.~\ref{ScLtVlt}).}  
\end{rmrk}
\section{Asymptotic for Cattaneo's prey-taxis}
\label{ScCttn}
\noindent
Here we make assumption  as in the case of PKS-model, and   also assume to hold true \emph{both} equalities in system~\eqref{LmtBhvrPrmtr} and  one more inequality, namely 
\begin{equation}\label{BrMuNtEqlCSqrd}
  c^2\neq \mu.
\end{equation}
%For applied models,  the diffusivity coefficients  often allow explicit upper and lower bounds uniform in $s>0$, which make the inequality  \eqref{BrMuNtEqlCSqrd}  easy to check.
 Equations \eqref{PrdTrnsEq}-\eqref{PryDnstEq}-\eqref{CttnFlxEq} written with regard to dependence on  the fast variable, $\eta$, read
 \begin{eqnarray}
&\left(q-cp\right)^\prime=\delta\left(pf(p,s)-p_{t}-q_{x}\right),&
\label{PrdTrnsEqAsmpt}\\
&\left(c{s}  +  {s}^\prime\right)^\prime=-\delta\left(sg(p,s)+\delta s_{xx}+2s^\prime_x-s_t\right),\quad&
\label{PryDnstEqAsmpt}\\
&\left(\mu p^\prime +\bar{\nu} q -cq^\prime -p\kappa  h^\prime-p\chi {s}^\prime\right)= \delta\left(p(\chi s_x+\kappa h_x) -\mu p_x-q_t\right).&
\label{CttnFlxEqAsmpt}
%\\
%&\tilde{s}=s-s_0=\delta s_1+\ldots,\ \bar{s}=s_0,&
%\label{StlSbr}
 \end{eqnarray}
 Replacing $p,q,s$ by expansions \eqref{AsmpExpnsn} leads to the following chain of equations 
\begin{eqnarray}
&\left(q_k-cp_k\right)^\prime=P_{k-1}-p_{k-1,t}-q_{k-1,x}+\mathrm{Op}_{k-1},&
\label{PrdTrnsEqChn}\\
&\left({\mu} p_k^\prime +\bar{\nu} q_k -cq_k^\prime -p_k{\kappa}  h^\prime-{\chi}(p_0s^\prime_k-p_ks_0^\prime) \right) =p_{k-1}(\chi s_{0,x}+\kappa h_x)+\chi p_0s_{k-1,x}-\mu p_{k-1,x}-q_{k-1,t}+\mathrm{Op}_{k-1}, &
\label{CttnFlxEqChn}\\
&\left(c{s}_k  +  {s}_k^\prime\right)^\prime=s_{k-1,t}-S_{k-1}-2s^\prime_{k-1,x}-s_{k-2,xx}+\mathrm{Op}_{k-1},&
\label{PryDnstEqChn}\\
&k=0,1,\ldots,&
\nonumber
 \end{eqnarray}
where all the quantities bearing negative  indices are equal to zero,   terms $P_{k-1}$,  $S_{k-1}$  are the same as in the case of PKS-model. 

For $k=0$,  Eqs.~ \eqref{PrdTrnsEqChn}-\eqref{PryDnstEqChn} become homogeneous in $s_0,p_0,q_0$, and the last one in effect reduces to $s_{0}^\prime=0$. Given this,  integrating the first one  in the fast variable and then eliminating  dependent variable $q_0$ from the  second one leads to equation
\begin{equation}\label{EqP0}
q_0=cp_0-c\bar{u}_0,\quad \left({\mu}-c^2\right) p_0^\prime +(\bar{\nu} c -{\kappa}  h^\prime) p_0  =\bar{\nu}c \bar{u}_0,
\end{equation}
where $\bar{u}_0$ is an unknown function in the slow variables. Here the  expression for $q_0$ is slightly different from   that given in Eq.~\eqref{P0Q0S0-PKS}. These changes   do not entail any losses in generality, and they are only to cover  the case of $c=0$ smoothly. Further, by integrating  Eq.~\eqref{EqP0}, we arrive at the following expressions 
\begin{equation}\label{Q0byP0AndP0byBarQ0}
q_0=cp_0-c\bar{u}_0,\quad p_0=\frac{b \bar{u}_0 \mathrm{R}_{b} F}{F},\ \ b=\frac{c\bar{\nu}}{{\mu}-c^2},\ \ F=\mathrm{e}^{\frac{\kappa \mathrm{R}_0h^\prime}{c^2-\mu}},\ s_0=\bar{s}_{0}(x,t).
\end{equation}
Here ${\mu}-c^2\neq 0$ by assumption \eqref{BrMuNtEqlCSqrd}, and slow functions $\bar{u}_0$, $\bar{s}_{0}$ remain undetermined.

The higher order slow system reads as
\begin{eqnarray}
&\left\langle P_{k-1} -p_{k-1,t}-q_{k-1,x}+\mathrm{Op}_{k-1}\right\rangle=0&
\label{PrdTrnsEqChnSlvCnd}\\
& \left\langle s_{k-1,t}-S_{k-1}-2s^\prime_{k-1,x}+\mathrm{Op}_{k-1}\right\rangle=0,&
\label{PryDnstEqChnSlvCnd}
%\\
%&\left({\mu} p_k^\prime +\bar{\nu} q_k -cq_k^\prime -p_k{\kappa}  h^\prime\right) ={\chi}p_0s^\prime_k+Q_{k-1}, &
%\label{CttnFlxEqChn}
 \end{eqnarray}
%\begin{equation}\label{EqP0}
%\left({\mu}-c^2\right) p_0^\prime +(\bar{\nu} c -{\kappa}  h^\prime) p_0  =\bar{\nu} \bar{q}_0,
%\end{equation}
Assume, we manage to resolve it. Then   Eqs.~\eqref{PrdTrnsEqChn} and~\eqref{PryDnstEqChn}
allow the integration,   which leads to equalities
\begin{equation}\label{QkSk}
  q_k=cp_k+\mathrm{R}_0(P_{k-1}-p_{k-1,t}-q_{k-1,x})-c\bar{u}_k,\quad s_k=\bar{s}_k+\mathrm{G}_c(s_{k-1,t}-S_{k-1}+\mathrm{Op}_{k-1})
\end{equation}
Next, we  eliminate variable $q_k$ from Eq.~\eqref{CttnFlxEqChn}  and arrive at the following equation
\begin{eqnarray}
&\left({\mu}-c^2\right) p_k^\prime +(\bar{\nu} c -{\kappa}  h^\prime) p_k  =c\bar{\nu} \bar{u}_k+ {\chi}p_0s^\prime_k+Q_{k-1}+\mathrm{Op}_{k-1},&
\label{EqPk}\\
& Q_{k-1}=p_{k-1}(\chi s_{0,x}+\kappa h_x)+\chi p_0s_{k-1,x}-\mu p_{k-1,x}-q_{k-1,t}-\bar{\nu} \mathrm{R}_0(P_{k-1}-p_{k-1,t}-q_{k-1,x})&
\nonumber
%\\
%&+\chi p_0 \mathrm{R}_c(s_{k-1,t}-S_{k-1}-2s^\prime_{k-1,x}),&
%\nonumber
\end{eqnarray}
By eliminating variable  $s_k$ by  Eq.~\eqref{QkSk}, we write down  the solution as follows
\begin{equation}\label{Pk}
  p_k=F^{-1}\mathrm{R}_bF(b\bar{u}_k+\mathrm{Op}_{k}),
\end{equation}
We now turn to the slow system  of $k-$the order, that reads exactly as system \eqref{PrdTrnsEqChnSlvCnd}-\eqref{PryDnstEqChnSlvCnd}, where index $k$ is increased by one. 
Substituting variables $q_k$, $p_k$ and $s_k$ from formulae \eqref{QkSk} and \eqref{Pk} brings the slow system of order $k$ into the form
\begin{eqnarray}
% \nonumber % Remove numbering (before each equation)
\bar{p}_{k,t}+\bar{q}_{kx}=\langle P_{k-1}+\mathrm{Op}_{k}\rangle,\ \ \bar{q}_{k}=\bar{c}\bar{p}_k+\langle \mathrm{Op}_{k}\rangle\ \ \bar{s}_{k,t}=\langle {S}_k +\mathrm{Op}_{k}\rangle, \ \ \bar{p}_k=\langle p_k\rangle,\quad \bar{c}=c\left(1-\left\langle \frac{F}{b\mathrm{R}_bF}\right\rangle\right).
\label{SlwSstGnrl}
\end{eqnarray}
The leading slow system takes the form
\begin{eqnarray}
&\bar{p}_{t}+\bar{q}_x=\bar{p}\bar{f}(\bar{p},\bar{s}),\quad \bar{q}=\bar{c}\bar{p}, \quad \bar{s}_t=\bar{s}\bar{g}(\bar{p},\bar{s}),&
\label{LdSlwSst}\\
& \bar{f}(\bar{p},\bar{s})=\langle\bar{F}f(\bar{p}\bar{F},\bar{s})\rangle,\quad \bar{g}(\bar{p},\bar{s})=\langle g(\bar{p}\bar{F},\bar{s})\rangle,\quad \bar{F}=\left\langle \frac{b\mathrm{R}_bF}{F}\right\rangle^{-1} \frac{b\mathrm{R}_bF}{F}&
\label{BrFBrGCttn}
\end{eqnarray}
where we have removed lower indices.

By applying the argument of Sec.~\ref{ScPKS} to the chain equations listed in this section, we conclude that  the asymptotic expansion \eqref{AsmpExpnsn} for Cattaneo's predator-prey system can be extended up to an arbitrarily assigned order. 
\begin{rmrk}\label{RmCttnSlwSst}
  {\rm The leading slow system derived here  regards purely the drift-kinetics interplay. It formally arises  from  the PKS slow systems  upon setting $\bar{\mu}=\bar{\chi}=0$, see Sec.~\ref{ScPKS}.} 
\end{rmrk}
\section{Linear stability of the quasi-equilibria}
\noindent\label{ScStQsEq}
Let the external short-wave signal  undergo no slow modulation -- that is,
\begin{equation}\label{NoSlwMdltn}
  h=h(\eta),\ \eta=x-ct.
\end{equation}
With such an assumption the slow system of every order gets the translational invariance in variable $x$ both in the PKS and in Cattaneo cases. Then it is of sense to consider 
the homogeneous equilibria of the leading slow system -- that is, the solutions such that 
  \begin{equation}\label{QsEqlbr}
 \bar{s}=\bar{s}_e\equiv\const,\quad\bar{p}=\bar{p}_e \equiv \const.
\end{equation}
We call the corresponding asymptotic solutions   quasi-equilibria. The quasi-equilibria patterns (if any) are  the short waves that propagate with no slow modulation at least in the leading approximation (see Eqs.~\eqref{P0Q0S0-PKS},~\eqref{Q0byP0AndP0byBarQ0}). We address their  stability  using the linear stability analysis for the equilibria of the leading slow system. 
\subsection{Quasi-equilibria}
\noindent
The quasi-equilibria densities, $\bar{p}_e$ and $\bar{s}_e$,  obey the following equations
\begin{equation}\label{BrPeSe}
  \bar{s}_e\bar{f}(\bar{s}_e,\bar{p}_e)=0,\quad\bar{s}_e\bar{g}(\bar{s}_e,\bar{p}_e)=0,
\end{equation}  
where formula \eqref{BrFBrGPKS} or \eqref{BrFBrGCttn} specifies functions $\bar{f}$ and $\bar{g}$ in  the PKS or Cattaneo's case. These expressions involve the exponential functions, $E$ and $F$, which we have assigned in Eqs.~\eqref{P0Q0S0-PKS} and~\eqref{Q0byP0AndP0byBarQ0}. In the subsequent work, it is convenient to unite the forms of writing them as follows
\begin{eqnarray}
 &\mathcal{E}=\mathrm{e}^{a h_0},\quad \text{where}& 
 \label{F-E-unfd}\\
 &a_0h_0=R_0h^\prime,\quad \langle{h_0^\prime}^2\rangle=1,&
 \label{hNrml}\\
 &a=\frac{\kappa a_0}{\mu},\quad \mathcal{E}=E,\text{ for the PKS-case or}&
 \label{a-PKS}\\
   & a=\frac{\kappa a_0 }{c^2-\mu},\ \mathcal{E}=F\text{ for Cattaneo' case.} &
 \label{a-Cttn}
\end{eqnarray}
Accordingly, we substitute the notation of  $\bar{E}$ ($\bar{F}$) in the PKS (Cattaneo's) case with
 \begin{eqnarray}
 &\bar{\mathcal{E}}=\mathcal{E}\langle\mathcal{E}\rangle^{-1}\ \text{ for the PKS-case or}& 
 \label{BrEPKSUnfd}\\
 &\bar{\mathcal{E}}=\frac{bR_b\mathcal{E}}{\mathcal{E}}\langle\frac{bR_b\mathcal{E}}{\mathcal{E}}\rangle^{-1}\ \text{ for Cattaneo' case.} &
 \label{BrFCttnUnfd}
\end{eqnarray}
Thus, the external signal takes the form $a_0h_0$ with amplitude $a_0$ and normalized profile $h_0$, while numbers $a$ defined in \eqref{a-PKS}-\eqref{a-Cttn} play the role of the effective amplitudes. Consequently, the equations  \eqref{BrPeSe} for the quasi-equilibria represent the integrals, which explicitly and smoothly depend on parameter $a$ (and  on $b$ as well in Cattaneo' case). Upon using the unified notation, they read 
\begin{equation}\label{EqnQsEq}
  \langle\bar{\mathcal{E}}\bar{p}f(\bar{\mathcal{E}}\bar{p},\bar{s})\rangle=0,\quad \langle\bar{s}g(\bar{\mathcal{E}}\bar{p},\bar{s})\rangle=0.
\end{equation}
\begin{lemma} \label{LmmQsiEqlbr}

{\sl Let the  original kinetics allow an non-degenerate equilibrium $p_e,s_e$ -- that is, the following equations   hold true 
\begin{equation}\label{Eqlbr}
  p_ef(p_e,s_e)=0,\quad s_eg(p_e,s_e)=0, 
\end{equation}
and the non-degeneracy condition}
\begin{equation}\label{NoDgnrcEqlbr}
0\neq\det\left.\left(
    \begin{array}{cc}
      a_{11} & a_{12} \\
      a_{21} & a_{22} \\
    \end{array}
  \right)\right|_{p_0=p_e,s_0=s_e},
\end{equation}
{\sl where Eq.~\eqref{aij} with $p_0=p_e, \ s_0=s_e$ specifies the matrix entries, $a_{ij}$.  
Then there exists a unique smooth 1-parametric family of the quasi-equilibria 
$\bar{p}_e=\bar{p}_e(a)$ and $\bar{s}_e=\bar{s}_e(a)$ defined in a neighbourhood of $0$ and such that $\bar{p}_e(0)=p_e$, $\bar{s}_e(0)=s_e$.}
\end{lemma}
{\bf Proof.} Consider equations~\eqref{EqnQsEq}  
relative to unknowns $\bar{p}_e,\bar{s}_e,a$. There is a solution
$$
\bar{p}_e=p_e,\quad \bar{s}_e=s_e,\quad a=0.
$$
The Jacoby matrix 
$$
\frac{\pr(p\bar{f}(p,s,a),s\bar{g}(p,s,a))}{\pr(p,s)}
$$
reads
$$
\left\langle \left(
    \begin{array}{cc}
      \mathcal{E}(f+p\mathcal{E}f_p)(p\mathcal{E},s) & p\mathcal{E}\bar{f}_s(p\mathcal{E},s) \\
      s\mathcal{E}\bar{g}_p(p\mathcal{E},s) & (s\bar{g})_s(p\mathcal{E},s) \\
    \end{array}
  \right)\right\rangle
$$
where the averaging acts on each entry. Since function $\mathcal{E}$ is smooth in  $a$, and $\left.\mathcal{E}\right|_{a=0}=1$, this Jacoby matrix is not degenerated for $a=0$ by assumption \eqref{NoDgnrcEqlbr}. Hence,  the lemma statement follows from the implicit function theorem. $\blacksquare$ 
\subsection{Linear stability analysis}
To examine the linear stability of the quasi-equilibria, we linearize the leading slow system near the quasi-equilibrium 
and look for a special solution  that reads as
\begin{equation}\label{EgnMds}
      (\hat{p},\hat{q},\hat{s})\exp(i k x+\lambda t),\ \hat{p},\hat{q},\hat{s},\lambda \in \mathbb{C},\quad k\in \mathbb{R},
\end{equation}
 It is common to name solutions \eqref{EgnMds} as the normal modes (of the corresponding equilibrium) and say that such a mode is \textsl{stable~(unstable, neutral)} if the real part of spectral parameter, $\lambda$,  is negative (positive, equal to zero). At that, the so-called spectral stability problem arises for which the spectral parameter $\lambda$ is the eigenvalue, and the corresponding normal mode is the eigenmode. It is also common to say that an equilibrium is stable~(unstable, neutral) provided every its normal mode is stable (there exists an unstable (neutral) mode).   Moreover,  the separation of the variables reduces the spectral stability problem to the algebraic eigenvalue one for every concrete wave number, $k$. 
 
Several authors (see Introduction for references)  had discovered occurrences of  instability for equilibria and onset of spatial-temporal patterns in the systems  dealt with here. They considered  various kinetics but  no external signal.  We, however, address the quasi-equilibria, and, consequently, move the focus on the instabilities due to the drift as it retains (together with the modified kinetics) all the memories of the external signal. 

The linear stability analysis for PKS-slow system  \eqref{PrdTrnsHmgnzdLdPKS}-\eqref{PryDnstHmgnzdLdPKS} leads to the following spectral stability problem 
\begin{equation}\label{MtrxStbGnrl}
 \det \left(
  \begin{array}{cc}
   a_{11} -({\hat{\mu}}+i\hat{c}+\lambda) & a_{12}+{\hat{\chi}}\\
   a_{21}  & a_{22}-\hat{\delta} \\
  \end{array}
\right)=0,
%{\bar{\chi}}_e={\bar{\chi}} \bar{p}_e,
\end{equation}
 where Eq.~\eqref{aij} determines coefficients $a_{ij}$ with $p_0=\bar{p}_e$, $s_0=\bar{s}_e$ and we have set
\begin{equation}\label{HatPrmtrs}
\hat{\delta}=k^2{{\delta}},\ \hat{\mu}=k^2{\bar{\mu}},\  \hat{\chi}=k^2{\bar{\chi}}\bar{p}_e,\ \hat{c}=k{\bar{c}}.
\end{equation}
Here, we replace equation \eqref{PryDnstHmgnzdLdPKS} by \eqref{PryDnstEq} for understanding  effect of the small prey diffusivity, with $D=\delta$, where parameter  $\delta$ is small positive. This spectral problem includes that arisen from   Cattaneo's slow system \eqref{LdSlwSst}  as a particular case with $\mu=0$ and $\chi=0$. 

We employ the method of Hankel's forms. It helps to count the eigenvalues with positive real parts. Namely, their number is equal to the number of the sign changes in the chain of Hankel' minors, which arises from the coefficients of the characteristic polynomials (we allude to  classical treatise \cite{GntMtrx} for details).

For a $2\times 2-$matrix the chain of Hankel's minors reads
 $$
 1,\Delta_2,\Delta_4,
 $$
where the subscript indicates the order of the minor.  For matrix~\eqref{MtrxStbGnrl}, we get
\begin{eqnarray}
  &\Delta_2=\hat{\mu}+\hat{\delta}-a_{{1,1}}-a_{{2,2}},\quad \Delta_4= \left({\hat{c}}^{2}+\Delta_2^2\right)(\hat{\delta} -a_{{2
,2}})\left(\hat{\mu}-a_{{1,1}} \right)-a_{{2,1}}\Delta_2^2(a_{1,2}+\hat{\chi})&
\label{HnklMnrsGnrl}
 \end{eqnarray}
From Eqs.~\eqref{HnklMnrsGnrl}, it follows that no instability occurs provided that
\begin{equation}\label{RstrAij}
a_{11}\le 0,\ a_{22}\le 0,\ a_{21}\le 0,\ a_{12}\ge 0
\end{equation}
where at least 3 inequality are strict. So, the drift does not effect the stability in this case. Restrictions \eqref{RstrAij}  makes sense, in particular they holds true for the Lotka-Volterra kinetics. We proceed with detailing this case in the next section. 
\begin{lemma}\label{LmmHnklChn}{\sl Let the parametric point 
$(\hat{\mu},\hat{\delta},\hat{\chi},a_{11},a_{12},a_{21},a_{22})$ 
satisfy  inequalities}
\begin{eqnarray}
&a_{11}a_{22}-a_{21}a_{12}>0,\ a_{11}+a_{22}<0,\  a_{11}a_{22}<0,&\ %a_{21}<0,
\label{InstLmmCnd1}\\
&(\hat{\delta} -a_{{2
,2}})\left(\hat{\mu}-a_{{1,1}} \right)<0,\ (\hat{\delta} -a_{{2
,2}})\left(\hat{\mu}-a_{{1,1}} \right)-a_{{2,1}}(a_{1,2}+\hat{\chi})>0,&
\label{InstLmmCnd2}\\
&\hat{\mu}\ge 0,\hat{\delta}\ge 0,\hat{\chi}\ge 0.&
\nonumber
\end{eqnarray}
{\sl Then, for every such a parametric point, there exists a threshold value, $\hat{c}^2_*>0$  such that the chain of signs of Hankel' minors, $1,\Delta_2,\Delta_4$, reads}
$$
\begin{array}{cc}
  +,+,+ \ \text{for}\ \hat{c}^2<\hat{c}^2_* \\
  +,+,-  \ \text{for }\ \hat{c}^2>\hat{c}^2_*.
\end{array}
$$
\end{lemma}
%
%(i) the homogeneous mode is stable  (ii) there exists a threshold value, $c^2_*>0$  such that  the unstable inhomogeneous ($k^2>0$) modes exist %for $\hat{c}^2>c^2_*$  and no such modes exist otherwise.
{\bf Proof.} Setting  to $0$ all the hat-terms puts  Hankel's chain into the following form
$$
1,\ -a_{11}-a_{22},\ a_{11}a_{22}-a_{21}a_{12}, 
$$ 
where all the elements are positive.  Furhter, inequalities \eqref{InstLmmCnd1}-\eqref{InstLmmCnd2} imply the positiveness for $\Delta_2$ for all the positive hat-terms. Next, isolating   $\hat{c}^2$ from equation $\Delta_4=0$ leads to an equality that 
reads
\begin{equation}\label{ThrsholdCSqrd}
  c_*^2=-\Delta_2^2\frac{\hat{\mu}\hat{\delta}-a_{2,1}\hat{\chi}-\hat{\delta}a_{1,1}-\hat{\mu}a_{2,2}+a_{1,1}a_{2,2}-a_{2,1}a_{1,2}}{(\hat{\delta} -a_{2,2})(\hat{\mu}-a_{1,1} )},
\end{equation}
where the right hand side iis positive by inequalities \eqref{InstLmmCnd1}-\eqref{InstLmmCnd2}. By the same inequalities, $\Delta_4$ is negative for $\hat{c}^2>\hat{c}^2_*$ and positive otherwise.   $\blacksquare$
 \begin{theorem}\label{ThInst}{\sl  Given the transport coefficients, $\mu,\chi,\delta,\kappa$, constant $c$ and function $h$ in one variable, which is smooth and periodic, consider a system of equations \eqref{PrdTrnsEq}-\eqref{PryDnstEq}  completed with the flux equation either in the form of PKS, \eqref{PKSFlxEq}, or Cattaneo's, \eqref{CttnFlxEq}. Let the external signal, $h$, take the form of a short travelling wave that propagate at the speed $c$ with no slow modulation -- that is, function $h$ meets condition  \eqref{NoSlwMdltn}. Assume the following
\begin{description}
  \item[(i)] the kinetics given allows a non-degenerated equilibria, $p=p_e,s=s_e$;
  \item[(ii)] all the inequalities \eqref{InstLmmCnd1} hold true for the coefficients, $a_{ij}$, defined by Eqs.~\eqref{aij} where  $p_0={p}_e$, $s_0={s}_e$.
 % \item[(iii)] there exists number $k_*>0$ such that all the inequalities \eqref{InstLmmCnd2} hold true for $\forall\, k^2\in (0,k_*^2)$ for %$\hat{\mu}=k^2\mu$, $\hat{\delta}=k^2\delta$,  $\hat{\chi}=k^2\chi$ and  $a_{ij}$ defined in clause (ii); 
%\item[(iiii)] if one of the  right hand sides in inequalities \eqref{InstLmmCnd2} gets to zero  on one of the  ends of the above interval, then it   satisfies  the non-degeneracy condition that  implies the embedding  of the corresponding end into a smooth branch of  zeroes of this right-hand side.
\end{description}}
{\sl Let $a\ge 0$ be effective amplitude defined in Eq.~\eqref{a-PKS}-\eqref{a-Cttn}. Then there exist $a_*>0$ and the branch of quasi-equilibria $\bar{p}_e=\bar{p}_e(a)$, $\bar{s}_e=\bar{s}_e(a)$,  which is smooth on $[0,a_*)$,  such that $\bar{p}_e(0)=p_e$,  $\bar{s}_e(0)={s}_e$, and for every  $a\in [0,a_*) $ there exist $k_*\in(0,\infty]$ (we do not exclude $k=+\infty$) such that for every $k:\,k^2\in(0,k^2_*)$ there exists $c_*\ge 0$ such that the normal mode linked to quasi-equilibrium $\bar{p}_e(a)$, $\bar{s}_e(a)$  with wave number $k:\, k^2\in(0,k^2_*)$ is stable  for $c^2<c^2_*$ and unstable for $c^2>c^2_*$ where $c$ is the speed at which the external signal propagates (see \eqref{NoSlwMdltn}).}
\end{theorem}
{\bf Proof.} Since the leading order system in fact does not depend on $\delta$, we shall  first prove the theorem for $\delta=0$.  In the course of proving, it will become   clear that the result persists for small $\delta$.  

Let $a$ be effective amplitude defined in Eq.~\eqref{a-PKS}-\eqref{a-Cttn}. By Lemma~\ref{LmmQsiEqlbr}, there exist the branch of quasi-equilibria $\bar{p}_e=\bar{p}_e(a)$, $\bar{s}_e=\bar{s}_e(a)$, which is smooth on $[0,a_*)$,  such that $\bar{p}_e(0)=p_e$,  $\bar{s}_e(0)={s}_e$, and and such that $\bar{p}_e(0)=p_e$,  $\bar{s}_e(0)={s}_e$, and coefficients, $a_{ij}$, are specified by  formulae \eqref{aij} with $\bar{p}_0=\bar{p}_e$, $s_0=\bar{s}_e$ satisfy  all the inequalities \eqref{InstLmmCnd1} for every $a\in[0,a_*)$. Hence, for every $a\in[0,a_*)$, there exists $k_*^2>0$ such that all the inequalities \eqref{InstLmmCnd2} hold true too upon expressing  the `hat'-parameters by formulae \eqref{HatPrmtrs}, where  $k^2\in[0,k_*^2)$. This is easy to see from inequalities \eqref{InstLmmCnd2} as 
  their  right hand sides turns out to be  linear  in variable $k^2$ (for $\delta=0$, of course), and both inequalities holds true for $k=0$ by assumption. Finally, we get the stability threshold, $c^2_*$, by Lemma~\ref{LmmHnklChn} with the use of  expression \eqref{ThrsholdCSqrd}.  $\blacksquare$ 
\begin{rmrk}\label{RmInstTrhsldExplct}
{\rm For the PKS-case (allowing small $\delta$), the threshold, $c^2_*$, reads
\begin{equation}\label{ThrhldC*Sqrd}
c^2_*=\frac{\Delta_2^2 M}{k^2(M-1)}\left(
\frac{a_{2,1}(k^2\bar{\chi}+a_{1,2})}{(k^2{\delta} -a_{2,2})(k^2\bar{\mu}-a_{1,1} )}-1\right),
\end{equation}
where $M=\langle\mathcal{E}\rangle\langle\mathcal{E}^{-1}\rangle>1$, $\bar{\mu}=\mu/M$, $\bar\chi=\chi \bar{p}_e/M$. Setting $\mu=\chi=0$ in this formula gives the threshold for Cattaneo's case. From Eq.~\eqref{ThrhldC*Sqrd}, it follows that the threshold goes to infinity when $k\to 0$, so that the homogeneous mode remains stable. It is consistent with the observation that setting $k=0$ entails the positiveness of all the minors in Hankel's chain under consideration. The effect of small $\delta$ reveals itself mainly by occurring the nearly infinite positive zero of the denominator in expression \eqref{ThrhldC*Sqrd}. This possibility does not  substantially change   the pictures outlined above and below as well.}
\end{rmrk}
\begin{rmrk}\label{RmInstPrprgt}
{\rm The instability described by this theorem is purely due to the propagating of external signal. Indeed, $M\to 1-0$, $a\to 0$, therefore the threshold \eqref{ThrhldC*Sqrd} goes to infinity. At the same time, if the external signal is steady, i.e.  $c=0$ and $\hat{c}=0$, therefore, then holding   inequalities \eqref{InstLmmCnd1}-\eqref{InstLmmCnd2}  true  is equivalent to the positiveness of all the minors in Hankel's chain under consideration. It entails  the stability of every mode the wave number of which obeys  inequalities \eqref{InstLmmCnd1}-\eqref{InstLmmCnd2} (together with the other parameters,  of course). Nevertheless, the possibility of exerting an effect remains open even for a steady signal provided that its amplitude is great enough. Such an occasion  is due to changing of the transport coefficients and the relative equilibria itself together with coefficients $a_{ij}$.  }
\end{rmrk}
\begin{rmrk}\label{RmEndPnts} {\rm  At least one of inequalities \eqref{InstLmmCnd2} must fail in the limit of $k^2\to k_*^2-0$. Let the other  one holds true as well as the all of inequalities \eqref{InstLmmCnd1}. Then the threshold for the external signal wave speed, $c_*^2$, tends to either $0$ or   $+\infty$.  The latter possibility arises from  violating the  first inequality of  set \eqref{InstLmmCnd2}. Then the normal modes with wave numbers $k:\,k^2>k_*^2$ get stable irrespective of the speed at which the external signal propagates, as Hankel's minor $\Delta_4$ becomes positive. Besides,  there is a positive lower bound for the values of $c_*^2$ attained for $k:\, k^2\in(0,k^2_*)$. This is not the case, if the former possibility  takes place,  i.e. the second inequality in set \eqref{InstLmmCnd2} gets violated. Moreover, the normal modes with wave numbers $k:\,k^2>k_*^2$ remains unstable, as Hankel's minor $\Delta_4$ remains negative while minor $\Delta_2$ remains positive. It means that the endpoints $c=0, k=\pm k_*$ of the neutral curve $c_*^2$ vs $k^2$ strike the neutral manifold of the other instability which is independent of the one under consideration, and which, in contrast, occurs in some domain on hyperplane  $c=0$.
}
\end{rmrk}
\begin{rmrk}\label{TtlStblt}
{\rm Theorem~\ref{ThInst} imply the stability for every mode with $k:\, k^2\in(0,k^2_*)$ provided that $c^2$ is smaller than the positive lower bound for $c_*^2$ (if any). This observation entails linear stability of the quasi-equilibria provided that $0<c^2<\inf\{c_*^2, k:\, k^2\in(0,k^2_*)\}$ and all the conditions of Theorem~\ref{ThInst} holds except for the first inequality \eqref{InstLmmCnd2}, perhaps. }
\end{rmrk}
\subsection{Lotka-Volterra's kinetics}  
\label{ScLtVlt}
Let's examine the quasi-equilibria  that arise from the Lotka-Volterra kinetics -- that is, we address  the particular cases of PKS and Cattaneo models that include equations \eqref{PrdTrnsEq} and \eqref{PryDnstEq} with 
\begin{equation}
\label{LtkVltrKntc}
f(p,s)=\gamma s-\beta,\quad g(p,s)=\alpha-p-s,\quad \alpha>0,\ \beta>0,\ \gamma>0,
\end{equation}
where $\alpha,\beta,\gamma$ are  the kinetic coefficients.  Bearing in mind that $\langle\bar{\mathcal{E}}\rangle=1$, we see that the Lotka-Volterra kinetics persists the homogenization. In other words,  functions \eqref{LtkVltrKntc} with the same values of the parameters  play the role of  the kinetic terms for the leading slow systems in both PKS and Cattaneo's cases. Furthermore, the quasi-equilibria densities $\bar{p}_e$ and $\bar{s}_e$ coincide with the equilibria ones, ${p}_e$and ${p}_e$, so that we use the notation with no  overlining in this subsection.  

The Lotka-Volterra kinetics allows the equilibrium of total extinction, $s_e=p_e=0$, and the equilibrium of the predators extinction, $p_e=0$, $s_e=\alpha$. The former is always unstable relative to every spatially homogeneous perturbation such that $s\neq 0$, and we'll not be considering it anymore. Additionally,  the equilibrium of coexistence becomes feasible, if 
\begin{equation}\label{PstvEqlbrCnstrnt}
  \alpha\gamma-\beta>0,
\end{equation}
and the corresponding densities read
\begin{equation}
\label{LtkVltrPstvEq}
s_e=\beta/\gamma>0,\ p_e=\alpha-s_e>0.
\end{equation}
Let's  examine  the stability for the quasi-equilibrium of coexistence in the framework of the leading slow system arisen from the PKS model. It consists of equations \eqref{PrdTrnsHmgnzdLdPKS}-\eqref{PryDnstHmgnzdLdPKS}. For understanding  effect of the small prey diffusivity we replace equation \eqref{PryDnstHmgnzdLdPKS} by \eqref{PryDnstEq} with $D=\delta$, where the diffusivity,  $\delta$, is small positive.

For the Lotka-Volterra kinetics, we have $a_{11}=0$, $a_{12}=\gamma p_e$, $a_{21}=a_{22}=-s_e$. These numbers do not meet condition~\eqref{InstLmmCnd1}. Hence,  we are not in a position to get any instabilities by Theorem~\ref{ThInst}.  At the same time, the inequalities \eqref{RstrAij} hold true. Hence, there are no unstable modes linked to the coexistence quasi-equilibrium irrespective of their wave numbers and other parameters the values of the other parameters provided that they allow it to exist. Furthermore, the inequalities  
\eqref{RstrAij} persist if we set ${\bar{\chi}}={\bar{\mu}}=0$. Since such a setting brings us at the spectral stability problem for  Cattaneo's case, we conclude  the quasi-equilibrium of coexistence is stable in the case of both Cattaneo's and the PKS prey-taxis  as long as it exists. 

%Let's check the stability directly. For this purpose, we replace coefficients $a_{ij}$ in formula \eqref{HnklMnrsGnrl} by their expressions listed above. After that the corresponding  Hankel's minors reads 
%\begin{eqnarray}
 % &\Delta_2=\hat{\mu}+\hat{\delta}+ s_e,\quad \Delta_4= \left({\hat{c}}^{2}+\Delta_2^2\right)(\hat{\delta}+s_e)\hat{\mu} +s_e\Delta_2^2(\gamma %p_e+\hat{\chi}).&
%\label{HnklMnrsLVCoExst}
 %\end{eqnarray}
%Formulae~\eqref{HnklMnrsLVCoExst} imply the stability for every mode. The same holds true for the leading slow system arising from Cattaneo's model. Indeed, all the minors remains positive for every admissible setting of the parameters including ${\bar{\chi}}={\bar{\mu}}=0$, $c=0$ and $\delta=0$. In particular, setting brings us.
%Hence, all the normal modes are stable for every wavenumber, and

Let's address the predators extinction equilibrium, $p_e=0$, $s_e=\alpha$, in the PKS framework.  This time  Hankel's chain reads
$$
1,\ ({\bar{\mu}}+\delta)k^2+\alpha+\beta_e,\ (k^2{\bar{\mu}}+\beta_e)(k^2\delta+\alpha)\left(c^2k^2+(({\bar{\mu}}+\delta)k^2+\alpha+\beta_e)^2\right)\quad \beta_e=\beta-\alpha\gamma
$$
There are no sign changes  for the wavenumbers, $k$, such that  $\beta_e+\bar{\mu}k^2>0$, and there is exactly  one sign change for every wave number such that $\beta_e+\bar{\mu}k^2<0$, and exactly one unstable normal mode, therefore. Hence, here we deal with a kind of long-wave instability. The pair of conjugated neutral modes exists for $k^2=-\beta_e/\bar{\mu}$. It is oscillatory in the sense that  $\mathrm{Im}\,\lambda\neq 0$ except $k=0$. The spatially homogeneous mode ($k=0$) gets unstable first, right when the value of $\beta_e$ gets negative, and   the corresponding bifurcation gives origin to the branch of the quasi-equilibria or equilibria of coexistence. 
\subsection{Ratio-dependent kinetics}
\label{ScRtDpnd}
\noindent
Consider the leading slow PKS and Cattaneo's systems with ratio-dependent  Arditi-Ginzburg kinetics. Upon a suitable scaling, it reads  
\begin{equation}\label{RtDpndKntc}
  f=\frac{\gamma\,s}{s+p}-\beta;\quad g=\alpha-s-\frac{rp}{s+p},\ r>0.
\end{equation}
where number $r$ plays the part of parameter.

Let the kinetic parameters, $\alpha,\beta,\gamma,r$, satisfy inequalities
\begin{equation}\label{RtDpndCnstrnt}
\gamma>\beta,\quad \gamma\alpha>r(\gamma -\beta).
\end{equation}
Then  the coexistence equilibrium reads
\begin{equation}\label{CxstnEqRtDpnd}
s_e=\alpha-r(1-\beta/\gamma);\quad   p_e=(\gamma/\beta-1)s_e.
\end{equation}
The conditions of Theorem~1 now read
%\begin{equation}\label{RtDpndCndSgnDet>0}
%\beta\gamma>r(\beta+\gamma), \quad \alpha\gamma^2<r(\gamma^2-\beta^2);
%\end{equation}
\begin{eqnarray}
&\beta<\gamma,\quad 0<r_3<r<\min(r_1,r_2)=\left\{\begin{array}{c}
                                     r_2, \gamma<\alpha+\beta \\
                                     r_1,  \gamma>\alpha+\beta
                                   \end{array}\right.
,\quad \text{where}&
\label{RtDpndPrmDmn}\\
& r_3=\frac{\alpha\gamma^2}{\gamma^2-\beta^2},\quad r_2=r_3+\frac{\gamma\beta}{\gamma+\beta},\quad r_1=r_3+\frac{\alpha\beta\gamma}{\gamma^2-\beta^2}.&
\nonumber
\end{eqnarray}
Thus, the conditions of Theorem~\ref{ThInst} simplifies to inequalities \eqref{RtDpndPrmDmn}, which obviously has nonempty solution.    Hence,  Theorem~\ref{ThInst} works and entails the instability for every mode such that $c_*^2(k^2)<c^2$. It remains true for $\delta=\bar{\mu}=\bar{\chi}=0$, and for the case of slow system arisen from Cattaneo's prey-taxis, therefore.
\section{Discussion}
\noindent\label{ScDscsn}
Thus, we have developed the power asymptotic expansions for the short travelling wave solutions for the PKS and Cattaneo's prey-taxis systems with short-wave external signal that itself represent a travelling wave propagating at speed $c$.  The important feature of the leading slow systems is the occurrence of drift, see Remark~\ref{RmLdSlwSstPKS}.  Morgulis at al. (2020) reported similar features for the leading slow system resulting from the short-wave asymptotic of the predator-prey system with the prey-taxis.
 For Cattaneo's prey taxis (see Morgulis (2023)) with no additional  assumptions on high resistivity (see \eqref{LmtBhvrPrmtr}) the short-waves contribute an additional  sources  into the leading slow flux systems. The assumption on high resistivity \eqref{LmtBhvrPrmtr} entails a very different asymptotic. 

For definiteness, let's proceed with the external signals with no slow modulation -- that is,  obeying constraint \eqref{NoSlwMdltn}. Then the drift velocity is constant, $\bar{c}$.  In the case of Cattaneo's prey taxis the leading slow flux equation simplifies to $\bar{q}=\bar{c}\bar{p}$,  and the leading  slow system is just a semi-linear hyperbolic one  written in Riemann's invariants  with characteristic velocities are $0$ and $\bar{c}$, and the higher order slow systems are linear hyperbolic with the same characteristic velocities.   

It worths noting, that $c-\bar{c}>0$ for both the PKS prey-taxis and for Cattaneo's one. In the former case it follows right by definition, see \eqref{BrCffcts} and \eqref{P0Q0S0-PKS}. In the latter case one also need to remind the definitions of operator  $R_z$ and parameters $a,b$ ( formulae~\eqref{Inv(D+z)} and~\eqref{Q0byP0AndP0byBarQ0}). 

One more corollary to the mentioned definitions is that 
$$
bR_b w\to \langle w \rangle\ \text{when}\ b\to 0\ \forall\, w\in \mathrm{L}_2(\mathbb{S}).
$$
Hence, $\bar{c}\to +0$ when $c\to 0$. Indeed,
$$
bR_bF\to   \langle E^{-1} \rangle,\ \left\langle \frac{b\mathrm{R}_bF}{F}\right\rangle^{-1}\to \frac{1}{\langle E^{-1} \rangle\langle E \rangle}.
$$
Thus, the drift occurs neither in the Cattaneo nor in the PKS model for $c=0$ --  that is, when the external signal is steady.  Such an observation contrasts to that reported for the indirect taxis by Morgulis \& Ilin (2020). Of course, switching off the signal immdeiately remove the drift and any modifications in the kinetics or transport coefficients.

Although the leading term of an asymptotic  gives us a summary of most substantial information of solution,  the high order approximations  can play important part   on the stage of justification. Regarding the asymptotics presented above, such a justification seems to be  feasible because only the leading slow systems are non-linear. 

We can produce the both  asymptotic expansions by solving a sequence of initial value problems for the leading slow systems. However, in the Cattaneo case the original system needs 3 initial functions, while the corresponding slow systems needs only 2 initial functions. This feature is natural as we are looking the solution in the form of short travelling wave, which can not fit general initial conditions.  Matching such a solution to the regular initial values   is likely requires a boundary layer near the initial line.  

Let's move focus on discussing the stability issues. The overall outcome of our study is that the short waves  is likely not capable of creating  the instability domain in the parametric space from nothing but it can substantially  widen the one that is non-empty with no signal. The details, however, essentially depend on the speed at which the external signal propagate, $c$, and on the system kinetics, at least,  when the signal is weak.    In particular, we have revealed the total stability total stability of the coexistence quasi-equilibria in the case of Lotka-Volterra's kinetic. This conclusion   agrees with what  Lee et al. have reported regarding this system with no external signal in the cited above paper. Indeed, the only difference between  the original PKS prey-taxis system with no external signal and the leading slow system is the term of $\bar{c}\bar{p}$ in equation \eqref{FlxHmgnzdLdPKS}. In other words, the systems coincides upon setting  $\bar{c}=0$ up to the scaling of the transport coefficients, $\mu$ and $\chi$ (see formulae~\eqref{BrCffcts}). 

In the case of Cattaneo' prey-taxis,  the leading slow system differs from the original system with no signal substantially. Besides, the latter allows the linear instability of the equilibrium of coexistence (see above cited article by Morgulis (2023)), which get unstable for  
$$
\chi>\chi_{cr}=\frac{\mu\nu+\delta C}{p_es_e},\ C=C(\nu,\gamma,p_e,s_e)>0.
$$
However, this result by no means contradicts to what we have been saying above regarding the stability domain as  we have derived the leading slow system  assuming $\nu\to\infty$ (see \eqref{LmtBhvrPrmtr}). Hence, the stabilization observed is due the infinite resistivity, $\nu$, rather than  to the external signal.

Let's get back to PKS prey-taxis system with the Arditi-Ginzburg kinetics (see Sec~\ref{ScRtDpnd} to remind the notation), and assume the external signal to be withdrawn. Then the coexistence equilibrium  can get unstable, nevertheless (see Lee at al., or  Banerjee \& Petrovskii ).   Generic neutral mode is either monotone (i.e. with $\IM\lambda=0$) or oscillatory (i.e. with $\IM\,\lambda\neq 0$). The latter one always has the complex conjugate counterpart. In particular, the values of $r_1$ and $r_2$ are critical in the sense that the coexistence equilibrium get unstable relative to spatially homogeneous mode, and the instability is monotone for $r=r_1$ and oscillatory for $r=r_2$. The former entails the switching from coexistence  to mutual extinction.

The spatially inhomogeneous modes can get unstable with no external signal too.  We allude to the cited above article  for the details regarding the PKS prey-taxis.   An  oscillatory unstable mode can occur for $r=r_{osc}>r_2$ only. In this sense, among the all of oscillatory modes the  homogeneous one always is the most unstable.  A monotone inhomogeneous mode can occur at some $r=r_{mnt}$, and  either $r_{mnt}<r_1$ or $r_{mnt}>r_1$ depending on its  wave length and the relations between the sensitivity, $\chi$, and diffusivities, $\delta$ and $\mu$. If $\delta=0$ (as in the leading slow system) the situation gets simpler: $r_{mnt}>r_1$ and the most unstable monotone mode is homogeneous provided that $\chi p_e>\mu$, where $p_e$ is the predator density in equilibria.  Otherwise, it is inhomogeneous and $r_{mnt}<r_1$.

Now let's get the external signal back and consider the quasi-equilibria for the leading slow PKS prey-taxis system.  Since its formulation presumes setting $\delta=0$,  we refrain from the considering $\delta>0$ (in fact, mostly for simplicity).  Additionally, let $\chi p_e>\mu$. Then no instability can occur except for that described by Theorem~\ref{ThInst}. Therefore, we assume that conditions \eqref{RtDpndPrmDmn} hold true and apply Theorem~\ref{ThInst}.  Then  Eq.~\eqref{ThrhldC*Sqrd} gives the  threshold value of the squared speed of the wave of the external signal, $c_*^2$.  It turns out that $c_*^2$ is smooth positive rational function in variable $k^2\in(0,\infty)$ -- that is, $k_*=\infty$, and  $c_*^2\to\infty$ for $k^2\to k^2_*=\infty $. Also, $c_*^2\to\infty$ for $k^2\to +0$. Hence,  $c^2_{cr}=\min\limits_{k^2>0}c_*^2>0$.  For $c:\, c^2<c^2_{cr}$ the coexistence quasi-equilibrium is stable. Otherwise, there exist unstable normal modes viz those having the wave numbers such that $c_*^2(k^2)<c^2$.  Hence, increasing the speed at which the wave of the external signal propagates exerts a destabilizing effect on the coexistence quasi-equilibrium provided the effective amplitude of the signal is sufficiently small. Of course, this smallness restriction does not depend on the wave speed, $c$.

 In the case of Cattaneo's prey-taxis, we can formally put  $\delta=\bar{\mu}=\bar{\chi}=0$ while analysing the stability of the quasi-equilibria. We assume conditions \eqref{RtDpndPrmDmn} to hold true again, and Theorem~\ref{ThInst} entails the instability for every normal mode such that $c_*^2(k^2)<c^2$. However,  this time  Eq.~\eqref{ThrhldC*Sqrd} gives $c_*^2=C^2/k^2$ for some $C>0$, so that $c_{cr}=0$. The result get more realistic if we  allow $\delta>0$. Then we can repeat the conclusions made for the parabolic case except for that regards the value of $k_*^2$. It is finite this time and $c^2_{cr}=\min\limits_{k_*^2>k^2>0}c_*^2>0$. A concrete   normal mode is unstable provided that $0<k^2<k_*^2$ and $c_*^2(k^2)<c$.  Hence, the destabilizing effect of increasing the propagation speed of the external signal persists upon passing to  the slow system emergent from the Cattaneo's prey-taxis model. It is surprising given the infinite resistivity  (in the sense of the asymptotic dealt with, see assumption \eqref{LmtBhvrPrmtr}).

The described cases both corresponds for the second option described in Remark~\ref{RmEndPnts}. The first option takes place for the parabolic slow system with  $\chi p_e<\mu$. Then, one can expect an interesting interplay of the unstable modes for small $c$ and $k$ such that $k^2$ is close to $k_*^2$. Here we mean the one induced by the external signal, and the other one is the monotone mode  occurring with no signal. 
\appendix
\section{Appendix}
\noindent 
We suppose the  transport coefficients to be assigned as they depend on the species under consideration, while the scales of the species densities,  predators flux, time and distance remain free to choose.  For the Cattaneo's prey-taxis, we choose the characteristic relaxation time as the time scale in order to normalize to the unity  the factor of $q_t$. For the parabolic prey-taxis, the time scale is the increment of the  prey density. Anyway, we  set a constraint $Q/X=P/T$ where $Q,P,X,T$ are the scales for flux, predators' density, distance and time, correspondingly. Then all the dimensional equation take the dimensionless form \eqref{PrdTrnsEq},\eqref{PryDnstEq}, \eqref{PKSFlxEq} and \eqref{CttnFlxEq} with reaction terms
$$
f(p,s)=T\widetilde{f}(Pp,Ss)\quad g(p,s)=Tsg(Pp,Ss), 
$$
where $S$ is the scale of the prey density, and $\widetilde{f}$, $\widetilde{g}$ are dimensional reaction terms. We can proceed further with normalizing the reaction terms by choosing suitable scales $T$, $P$ and $S$. In this way  we choose scales  $P,S$ that put the Lotka-Volterra kinetics into the form \eqref{LtkVltrKntc} and Arditi-Ginzburg kinetics into the form \eqref{RtDpndKntc}. Also,  we normalize $\alpha$ to 1 for the parabolic prey-taxis model by choosing scale $T$. We can normalize one more parameter by choosing $X$, but we do not do so, because the phenomena modelled can assign it naturally, for instance, as a  characteristic size of the lab equipment or biotop, or distance of migration. 
%%%%%%%%%%%%%%%%%%%%%%%%%%%%%%%%%%%%%%%%%%%%%%%%%%%%%%%%%%%%%%%%%%%%%%%%%%%%%%%%%%%%%%%%%%%%%%%%%%%%%
%%%%%%%%%%%%%%%%%%%%%%%%%%%%%%%%%%%%%%%%%%%%%%%%%%%%%%%%%%%%%%%%%%%%%%%%%%%%%%%%%%%%%%%%%%%%%%%%%%%%%%%%%%%%%
\end{document}